%% file: PaperConf.tex
\documentclass[10pt,conference]{IEEEtran}

\usepackage{color}

\usepackage{subfigure}

\usepackage[latin1]{inputenc}
\usepackage[pdftex]{graphicx}

\usepackage{amsmath}
\usepackage{amssymb}
\usepackage{amsfonts}

\IEEEoverridecommandlockouts

\begin{document}

\title{Multi-slot Coded ALOHA with Irregular Degree Distribution}



\author{
\authorblockN{Huyen-Chi Bui$^{1,2}$, Jérôme Lacan$^1$ and Marie-Laure Boucheret$^2$}
\authorblockA{$^1$University of Toulouse, ISAE/DMIA, \\
$^2$University of Toulouse, IRIT/ENSEEIHT\\
Email: \{huyen-chi.bui, jerome.lacan\}@isae.fr, marie-laure.boucheret@n7.fr
}
}


\maketitle

\begin{abstract}
This paper proposes an improvement of the random multiple access scheme for satellite communication named Multi-slot coded ALOHA (MuSCA). MuSCA is a generalization of Contention Resolution Diversity Slotted ALOHA (CRDSA). In this scheme, each user transmits several parts of a single codeword of an error correcting code instead of sending replicas. At the receiver level, the decoder collects all these parts and includes them in the decoding process even if they are interfered. In this paper, we show that a high throughput can be obtained by selecting variable code rates and user degrees according to a probability distribution. With an optimal irregular degree distribution, our system achieves a normalized throughput up to $1.43$, resulting in a significant gain compared to CRDSA and MuSCA. The spectral efficiency and the implementation issues of the scheme are also analyzed.
\end{abstract}


\input{part1_intro}

\input{part2_system_overview}

\input{part3_implementation}
\input{part4_perf}
\input{part5_conclusion}

\bibliographystyle{IEEEtran}
\bibliography{biblio}
\end{document}

%% file: part1_intro.tex
\section{Introduction}\label{introduction}

In a wireless communication system, two or more sources transmitting their data at the same time and on the same frequency generate interference. The interference between users are traditionally considered harmful. In first network generations, access methods strive to prevent simultaneous transmissions in order to avoid interference. Recently, an opposite approach named physical layer network coding (PNC) \cite{zhang:PNC} that allows interfered users to extract information from collided signals has motivated an extremely large number of studies. Instead of avoiding interference, PNC exploits it to increase system capacity \cite{Katti:PNC}. Asynchronous scenarios and practical deployment aspects of this technique have been studied \cite{Zhang:PNCasynchro, analogNC}.

Recent works have combined PNC with Successive Interference Cancellation (SIC) to resolve interference in the context of random access protocols. One of theses solutions is Contention Resolution-ALOHA (CRA) \cite{paperCRA} that transmits multiple replicas of a packet (called bursts) in a pure ALOHA system \cite{abramsonAloha}. In CRA scheme, users send randomly their bursts onto the communication medium. Each burst contains a signalling information which points to its replica locations. When a burst is successfully decoded, the replicas are also located and canceled. With a high Signal-to-Noise ratio (SNR) ($Es/N0 = 10$dB), the maximum normalized throughput is significantly extended from $T_{ALOHA} \approx 0.18$ up to $T_{CRA} \approx 0.98$. 

\begin{figure}[!t]
\centering
\includegraphics[width=6cm]{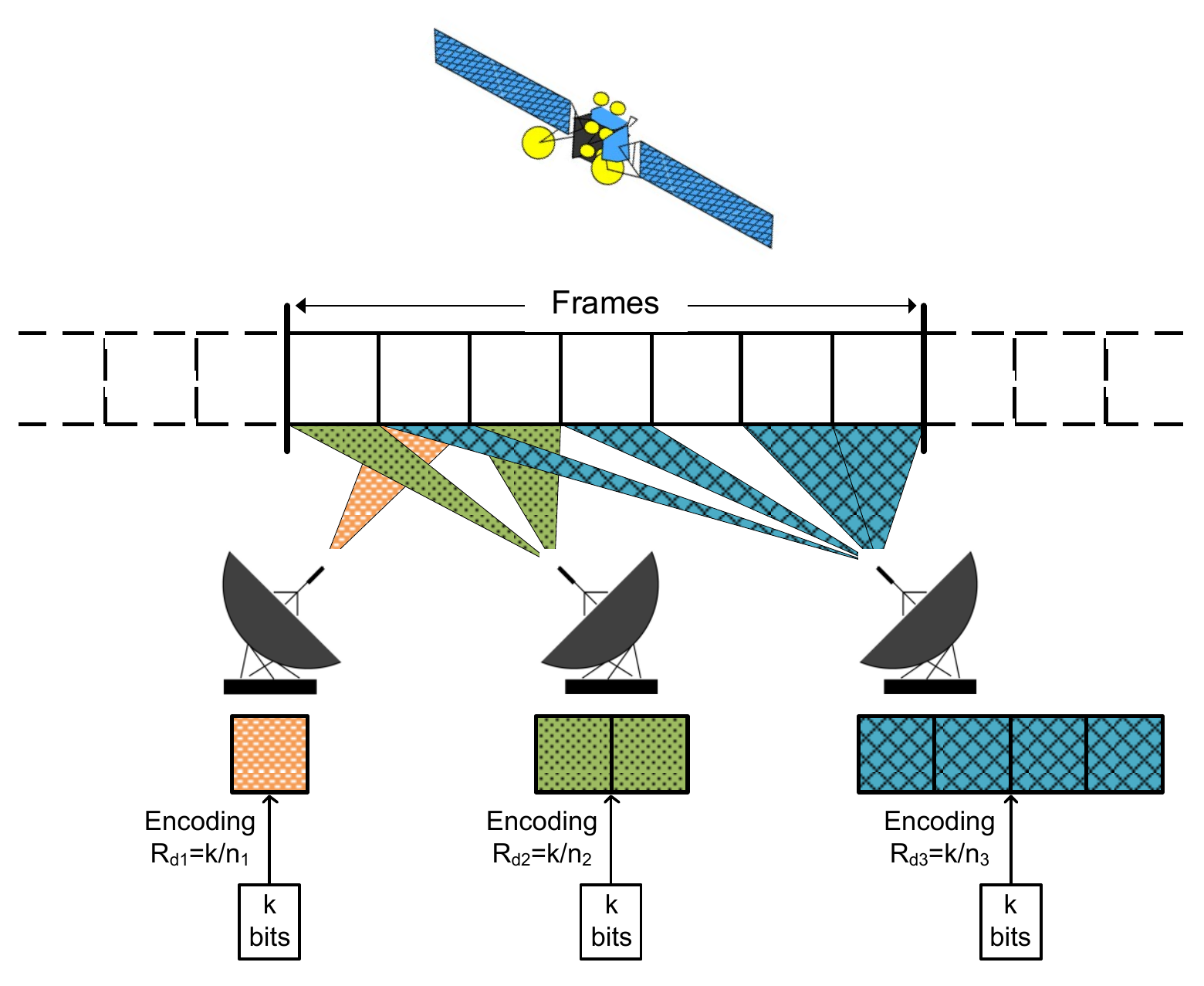}
\caption{Multiple access on a slotted channel}
\label{fig:system}
\end{figure}
In the same philosophy, PNC and SIC are applied to the Slotted ALOHA protocol \cite{paperSA} to create Contention Resolution Diversity Slotted ALOHA (CRDSA) and CRDSA++ (subsequently called CRDSA*) \cite{paperCRDSA, paperCRDSA++}. Unlike SA, instead of transmitting a single packet, every terminal additionally transmits one (CRDSA) or more (CRDSA++) replicas of the packet onto a frame of $N_s$ slots. As in CRA, the CRDSA* iterative decoding process is realized thanks to the header of each replica. The high performance of CRDSA* has motivated their adoption in the second generation of Digital Video Broadcasting Return Channel via Satellite (DVB-RCS2) standard \cite{DVB-RCS2-document}. Performance evaluations show that CRDSA* significantly outperforms SA in terms of normalized throughput ($T_{CRDSA} \approx 0.55 $, $T_{CRDSA++} \approx 0.68$ vs. $T_{SA} \approx 0.37$). 

A generalization of CRDSA, named Irregular Repetition Slotted ALOHA (IRSA) \cite{paperJournalIRSA}, proposes to apply variable repetition rates according to a probability distribution to each user. By optimizing the distribution, theoretical throughput can be increased up to $T_{IRSA} \approx$ 0.965 with a maximum repetition rate of 16. However, in a practical case where $N_s = 200$ slots, $T_{IRSA}$ is upper bound by 0.8. 

While CRDSA* and IRSA are based on repetitions, Coded Slotted ALOHA (CSA) encodes (instead of repeat) the bursts of each user with erasure correcting codes before the transmission \cite{paperCSA}. The maximum achievable throughput of CSA is $0.8$.

Recently, Multi-Slot Coded Aloha (MuSCA) \cite{paperMuSCA} was introduced as a new generalization of CRDSA. Instead of transmitting replicas, MuSCA sends several parts of a single codeword of an error correcting code. At the decoding level, the entity in charge of this process collects all bursts of the same user even if they are interfered. This is the main difference from CSA in which the decoder only considers non collided bursts. The decoder implements then a SIC process to remove successfully decoded signals. This random access method provides normalized throughput greater than 1.29 for a shot frame ($N_s = 100$ slots). This means a gain of $85\%$ and $75\%$ with respect to CRDSA++ with 3 repetitions (subsequently called CRDSA-3) and CSA.

Following the idea which generalized CRDSA to IRSA, we propose in this paper an optimization of MuSCA by applying variable code rates and irregular user degree distribution. The choice of the code rate of each user is done according to a probability distribution. The rest of the paper is organized as follows: Section~\ref{sec:Overview} provides an overview of the proposed scheme. Section~\ref{sec:impl} presents an implementation of our mechanism. The performance in terms of throughput is evaluated with simulations involving practical codes in Section~\ref{sec:evaluation}. Finally, we conclude the paper by summarizing the results and presenting the future work in Section~\ref{sec:Conclusion}.

%% file: part2_system_overview.tex
\section{System Overview}\label{sec:Overview}
\subsection{Definitions and Hypotheses}\label{sec:hypotheses}

We consider a wireless communication system with a satellite link shared among $N_u$ users (Figure~\ref{fig:system}). The satellite is a relay that amplifies all received signals with a fixed gain $G_r$. There is no direct link between users. The channel is considered linear and the transmission is subjected to Additive White Gaussian Noise (AWGN). We assume that each user has the same maximum energy per symbol $E_s$. The communication medium is divided into time and/or frequency slots of same size. We consider frames of duration $T_f$. $N_s$ consecutive slots form a frame, then the duration of each slot is $T_f/N_s$.

In our system, each user can attempt a transmit of $k$ information bits within a frame. To send more message, the user must wait until the next frame. We assume that the channel estimation is perfect and all users are synchronized at slot and frame levels. 

As in other wireless communication systems, if several users transmit their messages at the same time and frequency, there is collision. In this scheme, we integrate all bursts into the decoding process event if they are interfered by other bursts. 

\subsection{Description of the Mechanism}\label{sec:principle} 

As mentioned, each user transmits a data packet of $k$ bits per frame. The transmitter first encodes the packet with an error correcting code of rate $R_d$ and generates a codeword of $k/R_d$ bits. This codeword is modulated by a $M$-order modulator and split into $N_b$ part. The length of the data field of each part is thus $k/(R_d \times N_b\times log_2(M))$ symbols. Signalling information bits are added to each part to form a burst. The $N_b$ bursts from a data packet are then transmitted on $N_b$ random slots within a frame. In following sections, we call $N_b$ the user degree. $N_b$ and $R_d$ are selected according to the probability distribution defined by the system.

The relay receives a signal which is a noisy sum of the $N_b$ users signals after passing through the uplink channel. This sum can be written as
\begin{equation}
\label{sigRelay}
r_{relay}(t) = \sum \limits_{\substack{i=1}}^{N_u}c_i(t) s_i(t)+ n_u(t),
\end{equation}
where $s_i(t)$ is the signal transmitted by user $i$, $n_u(t)$ is the uplink AWGN with variance $\sigma_u = N_{0_u}/2$ and $c_i(t)$ takes into account the channel from terminal $i$ to the relay. The relay amplifies the input signal with a fixed gain $G_r$ and forwards this corrupted sum of messages back to all users on a second set of time slots or on another frequency.

The signal received by all users is
\begin{equation}
\label{sigReceiver}
r(t) = c_d(t) \times G_r \times \left(\sum \limits_{\substack{i=1}}^{N_u}c_i(t) s_i(t)+ n_u(t)) \right)+ n_d(t),
\end{equation}
where  $n_d(t)$ is the downlink AWGN with variance $\sigma_d = N_{0_d}/2$ and $c_d(t)$ takes into account the channel from the relay to the receiver. As in MuSCA with regular degree distribution, the receiver applies the SIC process twice to first decode signalling fields (to located bursts of each user) and then to decode data fields of located users. Once user $j$ is decoded, the receiver regenerates the signal of this user using estimated parameters (amplitude, frequency, phase) and subtracts it from the received signal. The resulting signal after the first round is given by
\begin{equation}
\label{sigReceiver1}
r_{1}(t) = c_d(t) \times G_r \times \left(\sum \limits_{\substack{i=1 \\ i\neq{j}}}^{N_u}c_i(t)s_i(t)+ n_u(t)) \right)+ n_d(t).
\end{equation}
The decoding algorithm is iterative until signal of all users are decoded or until a deadlock situation where no user is still decodable.

\subsection{Example of decoding algorithm}\label{sec:example}

\begin{figure}[!ht]
\begin{center}
\subfigure[]{
\includegraphics[keepaspectratio=true, width=0.52\columnwidth]{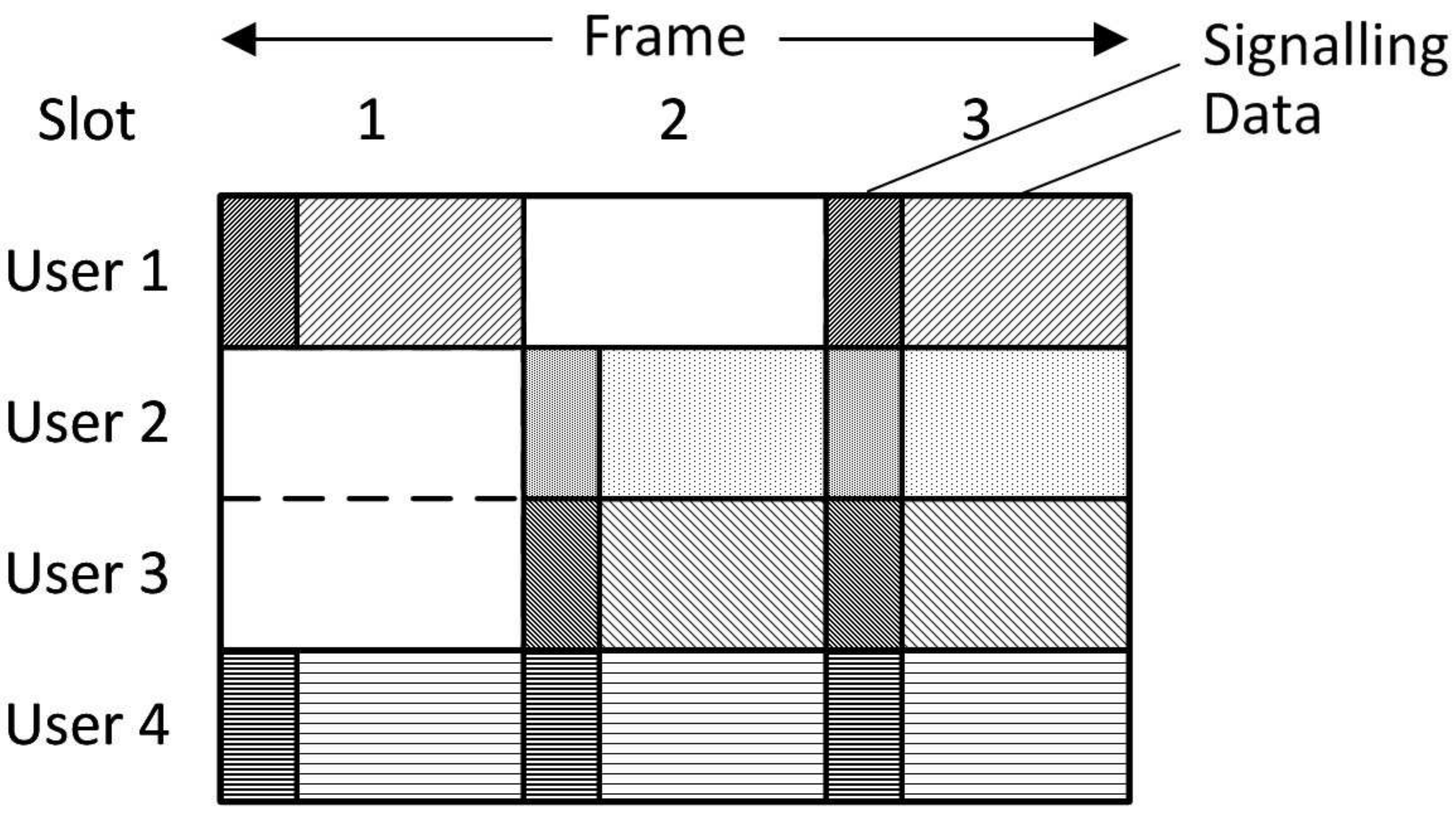}
\label{exempledecodeAlgorithm_a}
}

\subfigure[]{
\includegraphics[keepaspectratio=true, width=0.52\columnwidth]{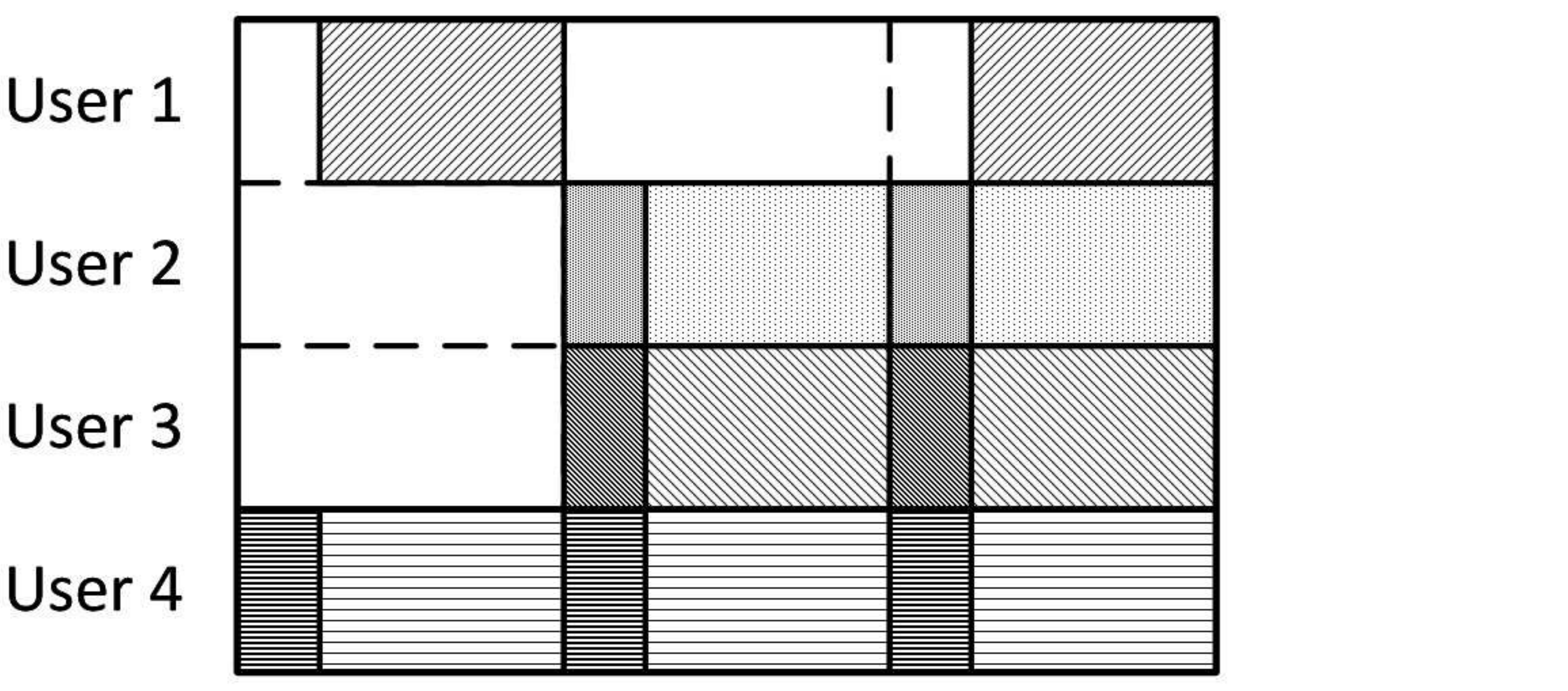}
\label{exempledecodeAlgorithm_b}
}

\subfigure[]{
\includegraphics[keepaspectratio=true, width=0.52\columnwidth]{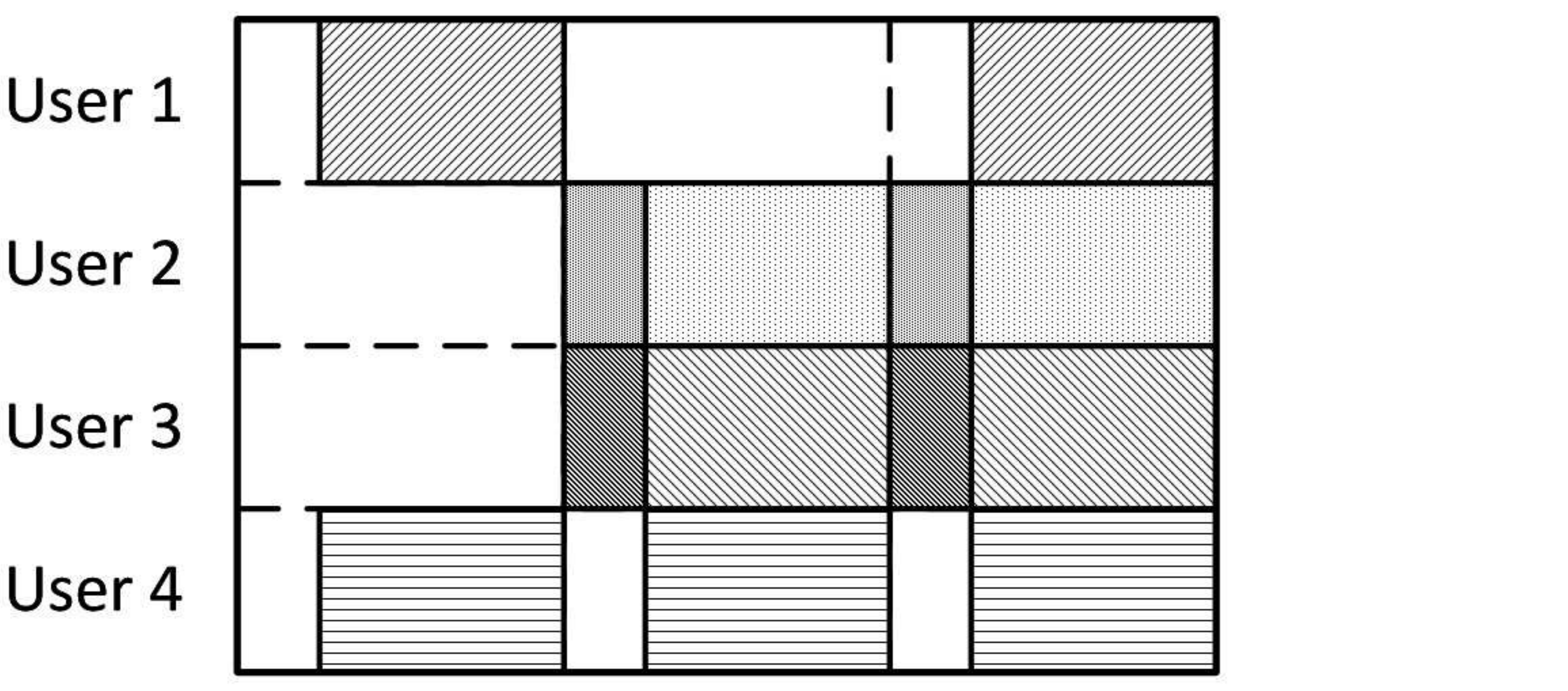}
\label{exempledecodeAlgorithm_c}
}

\caption{Example of signalling decoding}
\label{exempleDecodeSignalling}
\end{center}
\end{figure}

\begin{figure}[!ht]
\begin{center}

\subfigure[]{
\includegraphics[keepaspectratio=true, width=0.52\columnwidth]{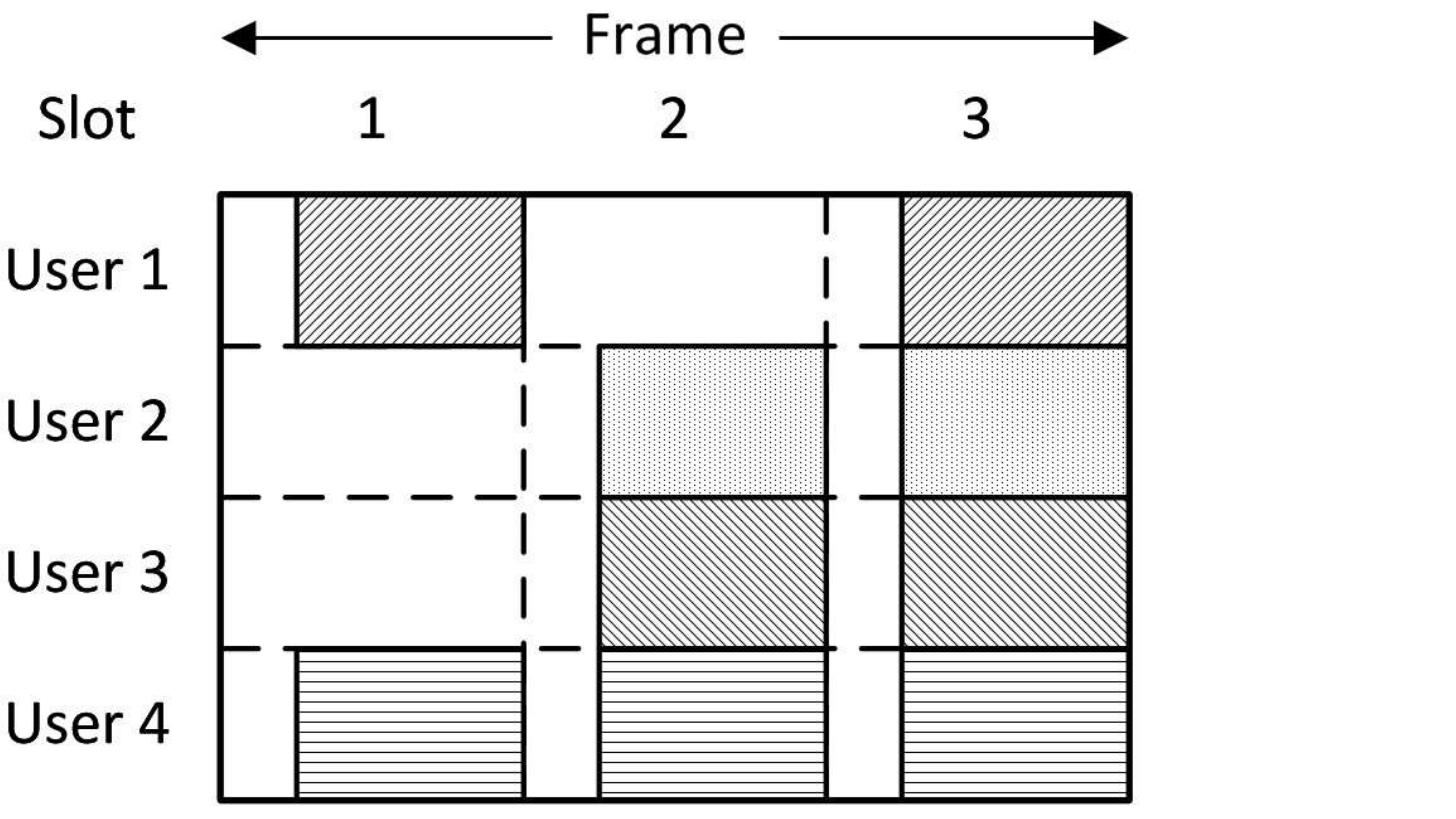}
\label{exempledecodeAlgorithm_d}
}

\subfigure[]{
\includegraphics[keepaspectratio=true, width=0.52\columnwidth]{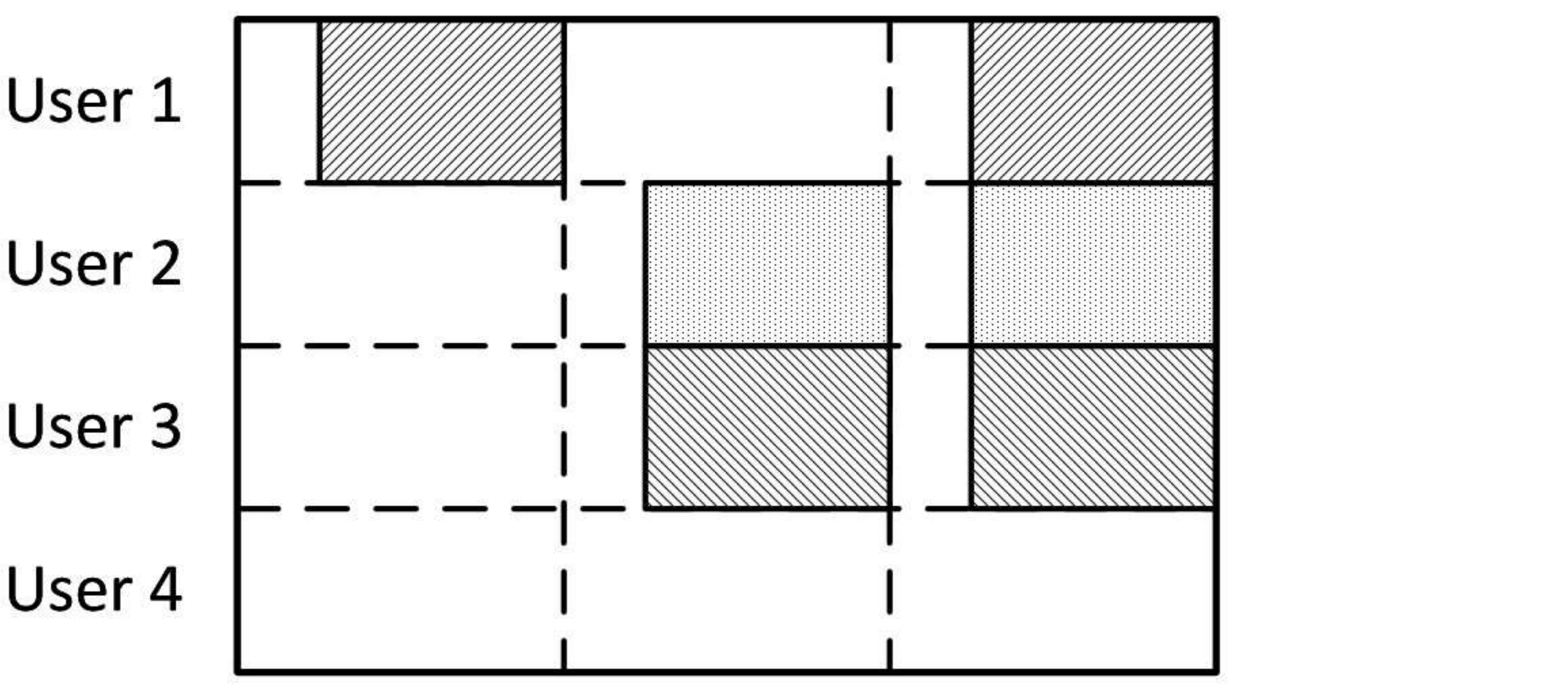}
\label{exempledecodeAlgorithm_e}
}

\subfigure[]{
\includegraphics[keepaspectratio=true, width=0.52\columnwidth]{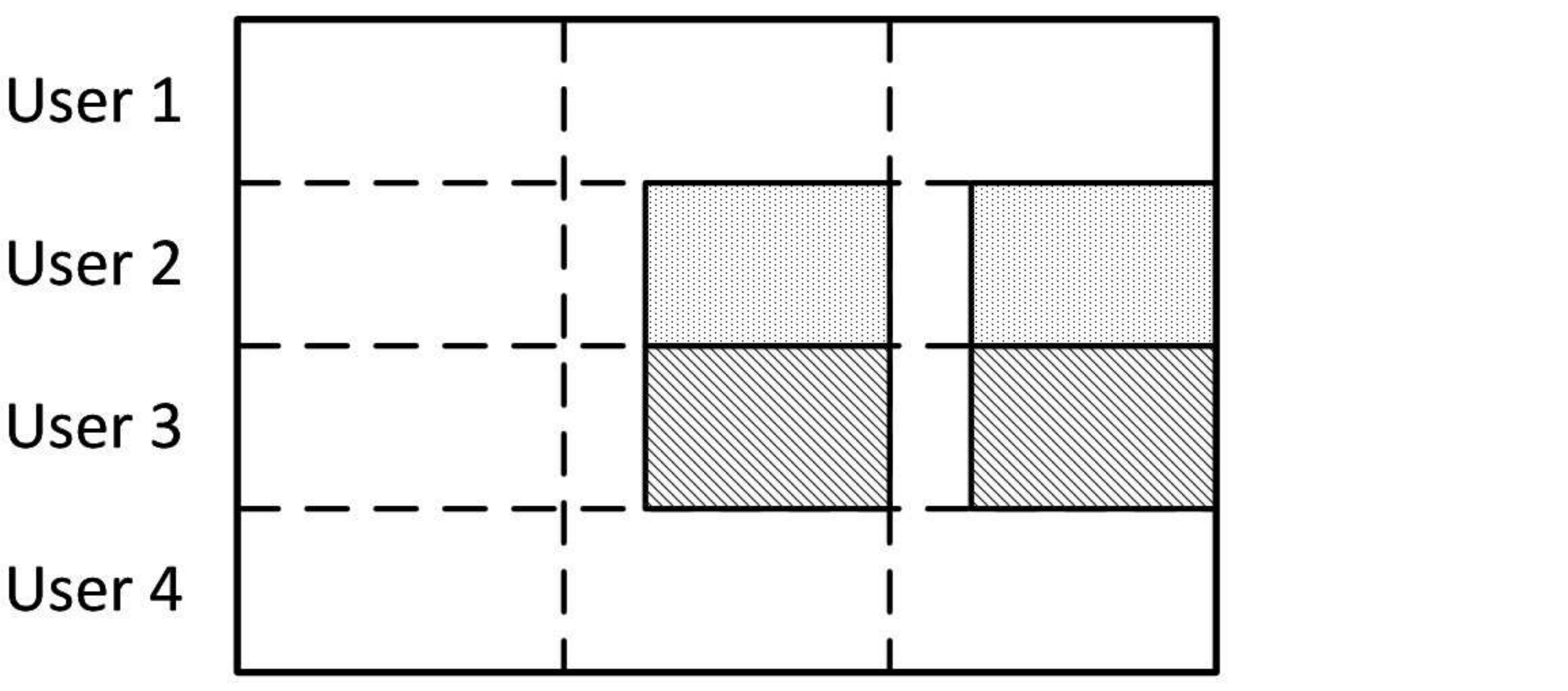}
\label{exempledecodeAlgorithm_f}
}

\caption{Example of data decoding}
\label{exempleDecodeData}
\end{center}
\end{figure}

Figure \ref{exempleDecodeSignalling} and \ref{exempleDecodeData} present an example of the decoding process. We consider a system with 4 users transmiting on a frame composed of 3 slots. The first 3 users use the same code of rate $R_{d1}$ to encode their packets of $k$ bits. The codeword of $n_1 = k/R_{d1}$ bits is divided in 2 parts ($N_b = 2$). The last user applies a code of rate $R_{d2}$ to its packet of $k$ bits. The codeword of $n_2 = k/R_{d2}$ is divided in 3 parts ($N_b= 3$). The code rates $R_{d1}$ and $R_{d2}$ are chosen in order to provide parts of same size (i.e., $R_{d1} = 3/2 R_{d2}$). In each part, pointers to other parts of the same user are added into the signalling field. This field is encoded by a short code of rate $R_s$. The performance of these 3 codes are known. Besides, we assume that the SNR is 5 dB for all users. The users send their bursts on the communication medium as represented in the Figure~\ref{exempledecodeAlgorithm_a}. Note that all bursts are collided with others. Therefore, this situation does not allow algorithms like CRDSA, IRSA and CSA to decode while our scheme is able to implement the decoding even if there is no clean burst on the frame.

The first phase concerns the decoding of signalling fields. The entity that handles the decoding runs through the entire frame to find the most likely decodable fields (in a collision-free slot or interfered by only one other user). Thus, on the slot 1, the decoder detects bursts that are interfered by one user, so it begins by decoding the first slot. With the code considered in Section~\ref{subsec:signalling}, simulation shows that the probability of failed decoding is 0.109. This signalling field has a high chance (90\%) to be correctly decoded. We suppose that the decoding is successful and the signalling field of user 1 is obtained. Then, the pointer to the second burst of the first user is recovered. The signalling fields of the both bursts are then subtracted from the received signal (see Figure~\ref{exempledecodeAlgorithm_b}).

Currently, the signalling field of the clean burst on slot 1 of user 4 is easily decodable. Its complementary bursts are located and their signalling fields are subtracted (Figure~\ref{exempledecodeAlgorithm_c}).

On slots 2 and 3, signalling fields of users 2 and 3 are interfered by 1 user. As before, signalling fields on slot 2 have a probability of $90\%$ to be decoded. Supposing that we obtain the pointers of user 2 on the slot 2, then we can subtract the both signalling fields of this user. After that, the signalling field of user 3 is clean (not in collision with other burst). The probability of locating the both bursts of this user is greater than $1-(10^{-4})^2$.

At the end of the first phase, the decoder has information on the location of all users' bursts. It starts the data decoding phase. User 4 has the highest decoding probability. Its 3 bursts are interfered with 1, 2 and 3 other bursts, respectively (Figure~\ref{exempledecodeAlgorithm_d}). We note this configuration [1~2~3]. The decoder starts the data decoding process with user 4. It gathers the 3 located bursts of this user to attempt the decoding. Simulations show that with the code considered in Section~\ref{subsec:data}, the packet error rate is $PER_{R_{d2}, [1 2 3]}=0.02$ at 5 dB. Supposing this user is successfully decoded, its 3 bursts can be regenerated and subtracted from the signal received in the corresponding slots (i.e. slots 1, 2 and 3) (Figure~\ref{exempledecodeAlgorithm_e}). User 1 becomes the one with highest decoding probability, $PER_{R_{d1}, [0 2]} < 10^{-4} $. After the successful decoding of user 1 (Figure~\ref{exempledecodeAlgorithm_f}), user 2 and 3 have the same decoding probability ($PER_{R_{d1}, [1 1]} < 10^{-4}$). We assume that user 2 is decoded first. The remaining user (user 3) is not any more interfered on the both slots and then easily decoded.

%% file: part3_implementation.tex
\section{Implementation}\label{sec:impl}

In \cite{paperMuSCA}, we showed that the choice of codes for the signalization field, payload and the parameters $N_b$, $N_s$ may deeply influence the system performance. In our scheme, each user could have different code and different degree from others. 

\subsection{Distribution probability}\label{subsec:distribution}

Our system works as follows: for each transmission, the user adopts a variable code rate $R_d$ and a degree $N_b$, which are selected according to a given distribution $\Lambda$. The distribution has to be optimized to increase the system performance in terms of throughput. Note that it is possible to choose independently $N_b$ and $R_d$. However, in this paper, we only detail the case where $R_d$ depends on $N_b$ (i.e., each codeword contains a fix amount of information bits). The distribution can be then represented as
\begin{equation}
\label{equDistibution}
\Lambda(x) = \sum \limits_{\substack{N_{b_i}}} P(N_{b_i}) \times x^{N_{b_i}},
\end{equation}
where $P(N_{b_i})$ is the probability that a user has the degree $N_{b_i}$ and $\sum{P(N_{b_i})} = 1$. 

\subsection{Signalling field}\label{subsec:signalling}
\begin{figure}[!t]
\centering
\includegraphics[width=0.9\columnwidth]{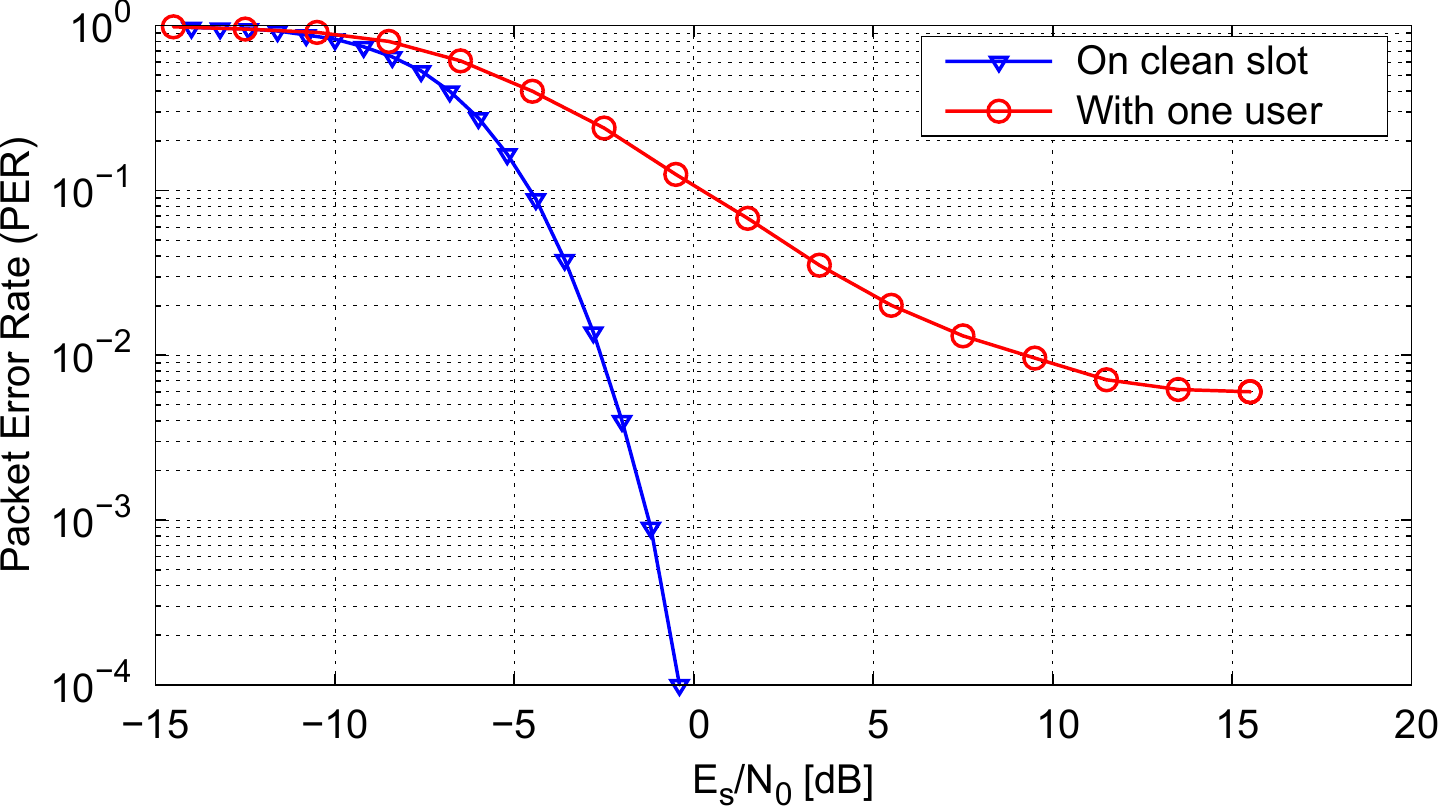}
\caption{Reed-Muller code (14, 64), BPSK modulation}
\label{PerformanceReedMuller14by64}
\end{figure}

As in regular MuSCA, the signalling field includes $N_b~-~1$ pointers to the positions of other bursts of the same user. Pointer sizes depend on the frame length $N_s$ and the code rate $R_s$. The signalling field size of each user can be estimated by
\begin{equation}
\label{equLengthSF}
L_s = \left\lceil log_2(N_s)\right\rceil \times R_s \times (N_b - 1).
\end{equation}
$L_s$ must remain relatively small in order to maintain a high throughput. We choose to protect these fields by Reed-Muller codes which have a fair performance/complexity trade-off for soft-decision decoding of short codewords \cite{paperRM}. For example, for a frame of length $N_s = 100$ slots, each location is represented by 7 bits. Users of degree 3 can adopt a  Reed-Muller code (14, 64). The signalization decoding is launched when a burst is on a clean (without collision) slot or it is interfered by only one user. Figure \ref{PerformanceReedMuller14by64} represents the PER curves of this code combined with BPSK modulation in 2 cases: the burst on a clean slot and the burst interfered with one user.

\subsection{Data field}\label{subsec:data}

Our scheme has no particular constraint on the choice of codes to encode data packets for each user. However, to be comparable with existing methods, we consider codes allowing to sent an amount of information bits per slot equivalent to CRDSA*. In CRDSA*, a source transmits $N_b$ replicas of the same packet within a frame. Each data packet is coded by a convolution code \cite{paperCRDSA} or turbo code \cite{paperCRDSA++} of rate $r = 1/2$. In our scheme, this code rate is equivalent to a general code of rate $R_d = 1/(2N_b)$ where $N_b$ depends on a irregular degree distribution $\Lambda$. Moreover, due to the system characteristic, the correcting codes have to simultaneously manage errors and collisions. In our simulations, we use CCSDS turbo codes \cite{CCSDSref} provided by the CML library \cite{libCML}. For $N_b = 1$, 2 or 3, the code rate $R_d$ is $1/2$, $1/4$ and $1/6$, respectively. Turbo codes, associated with QPSK modulation are applied to information bit sequences of length $k = 456$, producing codewords of $456/2R_d$ symbols. Note that the turbo codes do not reach their highest performance when there are long damaged sequences in the received codeword. Therefore, similarly to Digital Video Broadcasting - Satellite Handheld (DVB-SH) \cite{DVBSH-guidelines}, we apply a bit-interleaver to each codeword. 

\begin{figure}[!t]
\centering
\includegraphics[width=0.9\columnwidth]{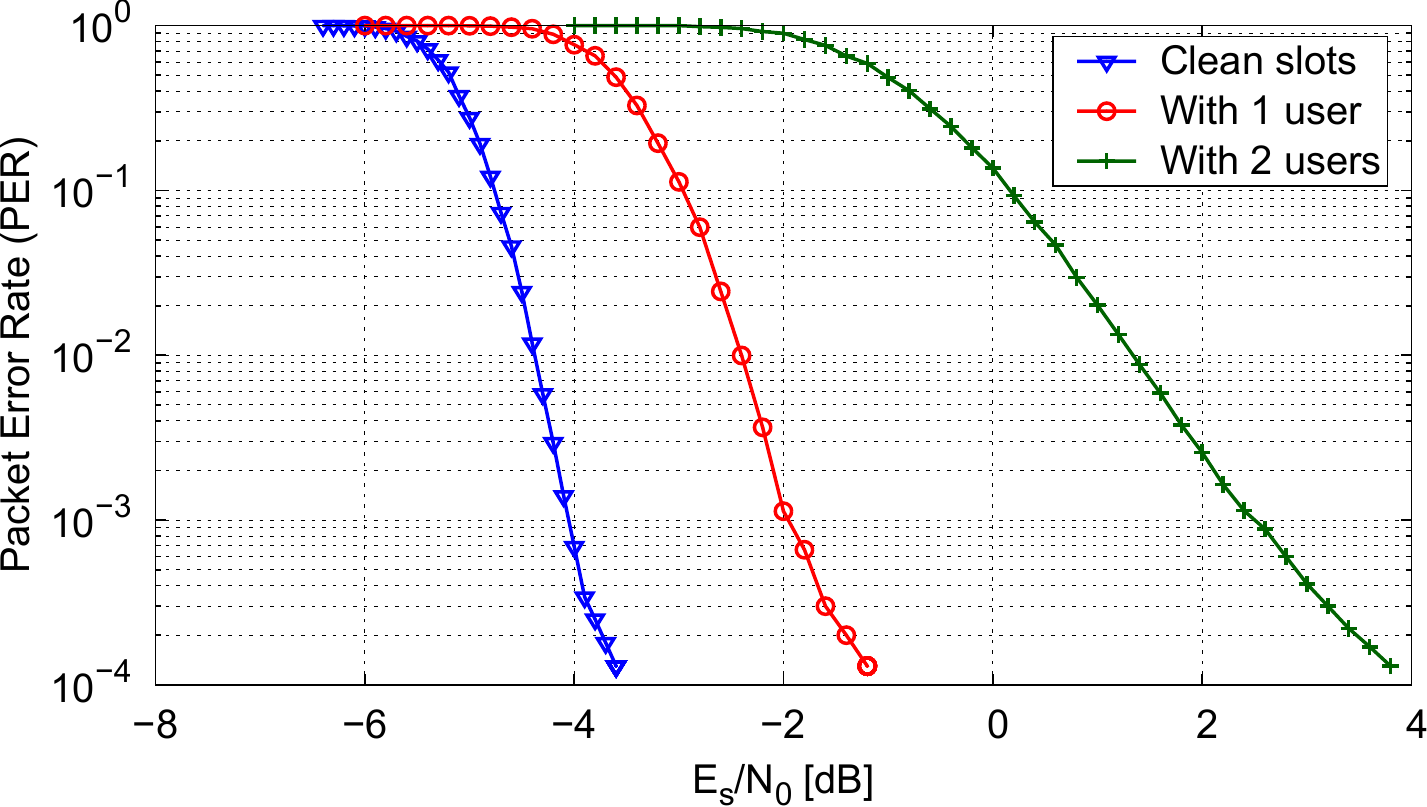}
\caption{Turbo code of $R_d$ = 1/6, k = 456, QPSK modulation, AWGN channel}
\label{PerfTurbo1by6_456}
\end{figure}

Figure \ref{PerfTurbo1by6_456} depicts the performance curves in terms of PER of a turbo code $R_d = 1/6$ combining with QPSK modulation in the 3 cases:

\begin{enumerate}
\item 3 bursts on clean slots (configuration [0~0~0]);
\item 3 bursts in collision with signal of one user of the same power (configuration [1~1~1]);
\item 3 bursts in collision with two other users (configuration [2~2~2]).
\end{enumerate}  
Bursts interfered by more than 2 other bursts are considered as erased. 

%% file: part4_perf.tex
\section{Performance evaluation}\label{sec:evaluation}

As mentioned in the previous section, we consider a frame composed of $N_s$ slots in which $N_u$ users attempt to transmit a data packet. The normalized load $G$ represents the average number of packet transmissions per slot and is defined by

\begin{equation} 
\label{normalizedLoad}
G = \frac{N_u}{N_s}.
\end{equation}

For each value of $E_s/N_0$ and $G$, we obtain a probability of non decoding a packet, denoted $PLR$. For a fixed $E_s/N_0$, the normalized throughput $T$ (defined as the number of successful packets transmissions per slot) is computed as

\begin{equation} 
\label{normalizedThroughput}
T = G \times {(1 - PLR(G))}.
\end{equation}

For SA, the throughput depends on $G$ as $T(G) = G \times e^{-G}$. SA is a particular case of our scheme where the distribution is $\Lambda_1(x)=x$. Besides, regular MuSCA with $N_b = 3$ (called MuSCA-3) corresponds to the distribution $\Lambda_3(x) = x^3$ associated with turbo codes.

In previous work, we showed that MuSCA can achieve a normalized throughput up to 1.29 \cite{paperMuSCA}. This means that, on average, more than one user can transmit their data packets per slot. Section \ref{sec:example} is an example where the normalized throughput is beyond 1. This throughput shows a significant gain between MuSCA and other presented methods. As in regular MuSCA, a frame of 100 slots is considered for all simulations in this study. A longer frame could improve the throughput but it increases the transmission delay because the decoding process cannot be started before the end of entire frame. 

In Section~\ref{subsec:signalling}, the link between the signalling field size $L_s$ and the user degree $N_b$ was presented. With a higher user degree (i.e., more pointers in a burst header), the traffic part occupied by the signalization will be larger compared to the useful signal. Therefore, we choose to limit the user degrees to 3 in order to maintain a high effective throughput.

\begin{table}[ht]
\begin{center}
\footnotesize{
\begin{tabular}{|c|c|}
\hline
Distribution $\Lambda(x)$	& Throughput $T$\\
\hline
$\Lambda_1(x) = x$   & 0.368\\
\hline
$\Lambda_2(x) = x^2$   & 1.270\\
\hline
$\Lambda_3(x) = x^3$   & 1.293\\
\hline
$\Lambda_4(x) = 0.7x^2 + 0.3x^3$   & 1.401\\
\hline
$\Lambda_5(x) = 0.1x + 0.3x^2 + 0.6x^3$& 1.426\\
\hline
$\Lambda_6(x) = 0.2x + 0.3x^2 + 0.5x^3$& 1.366\\
\hline
\end{tabular}}
\end{center}
\caption{Throughput computed for various distributions, $E_s/N_0 = 8$ dB}
\label{tab:maxT}
\end{table}

Theoretical evaluation based on density evolution (DE) \cite{paperJournalIRSA, paperCSA} is not applied in this study because DE relies on the hypothesis of the independence between exchanged messages in the iterative decoding process. This hypothesis is not verified for the short frame considered in our study. Moreover, this method does not allow to evaluate distributions that contain user of degree 1 while this degree is used in our optimal distribution. Therefore, in order to point out the optimal distribution, we performed simulations in varying the probabilities of each degree. As mentioned in Section \ref{subsec:data}, all users apply turbo codes to encoded packets of 456 bits ($k = 456$). Table~\ref{tab:maxT} represents maximum normalized throughputs for various degree distributions at a high SNR level ($E_s/N_0 = 8$dB). The optimal distribution obtained by simulations is $\Lambda_5(x) = 0.1x + 0.3x^2 + 0.6x^3$. The corresponding threshold exceeds 1.42 for a frame of length 100 slots. This is equivalent to the performance of regular MuSCA with a frame of 500 slots \cite{paperMuSCA}.

\begin{figure}[!ht]
\centering
\includegraphics[width=0.9\columnwidth]{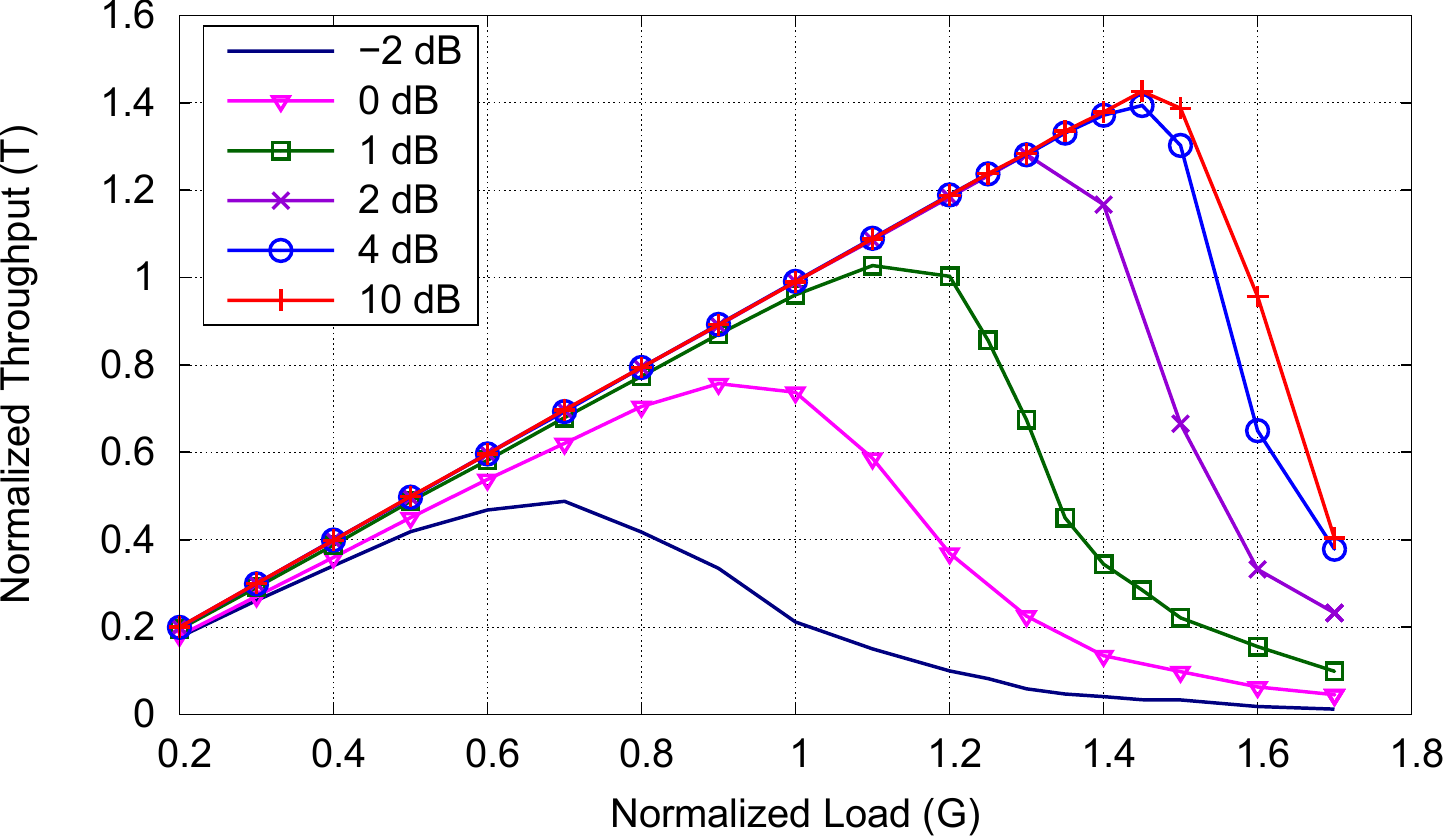}
\caption{Normalized throughput of MuSCA with optimal degree distribution $\Lambda_5(x)$ for various values of SNR} 
\label{fig:OptimalDistribution}
\end{figure}

\begin{figure}[!ht]
\centering
\subfigure[Normalized throughput ($T$)]{
\includegraphics[keepaspectratio=true, width=0.9\columnwidth]{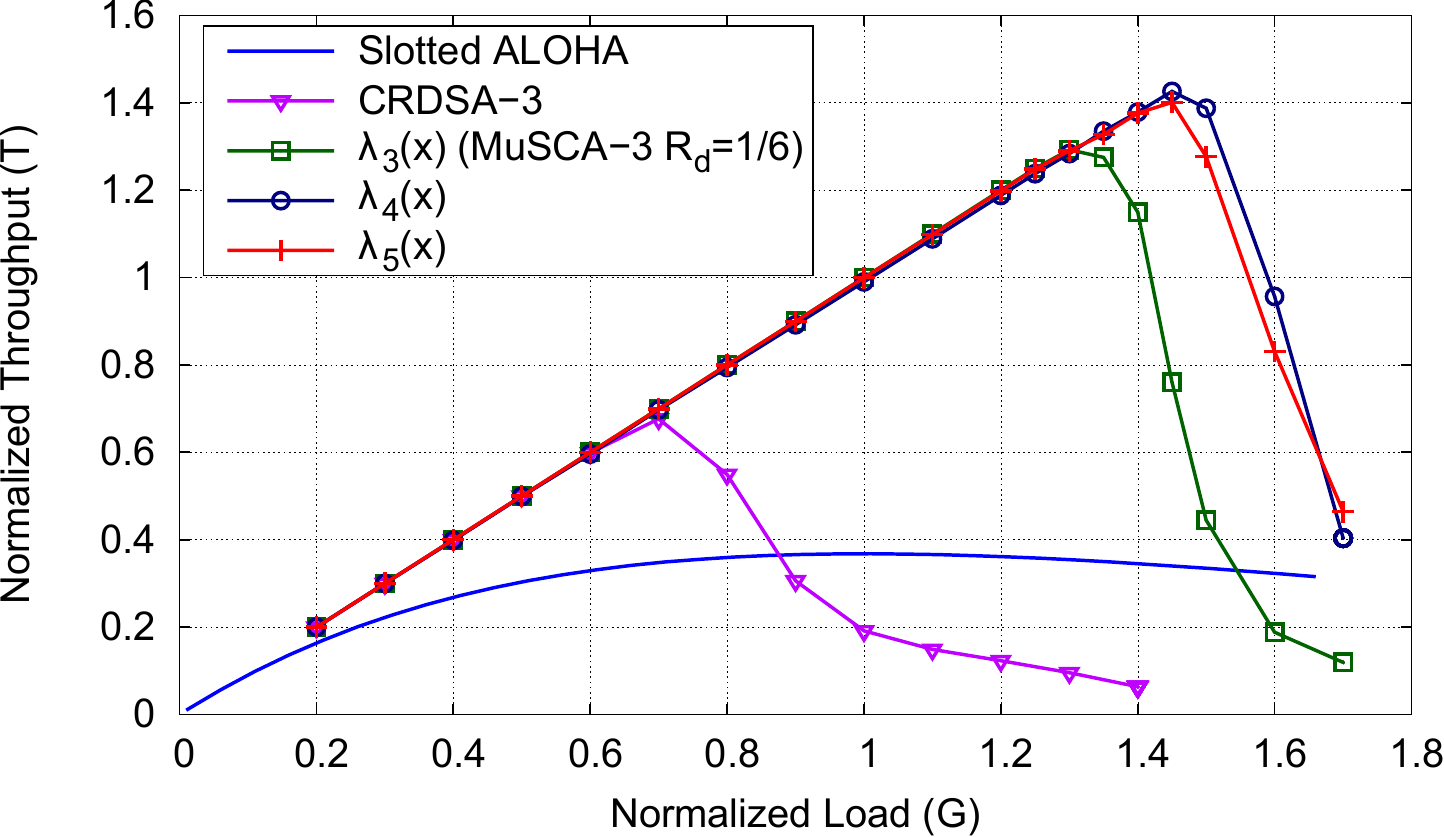}
\label{figT}
}
\subfigure[Packet loss ratio ($PLR$)]{
\includegraphics[keepaspectratio=true, width=0.9\columnwidth]{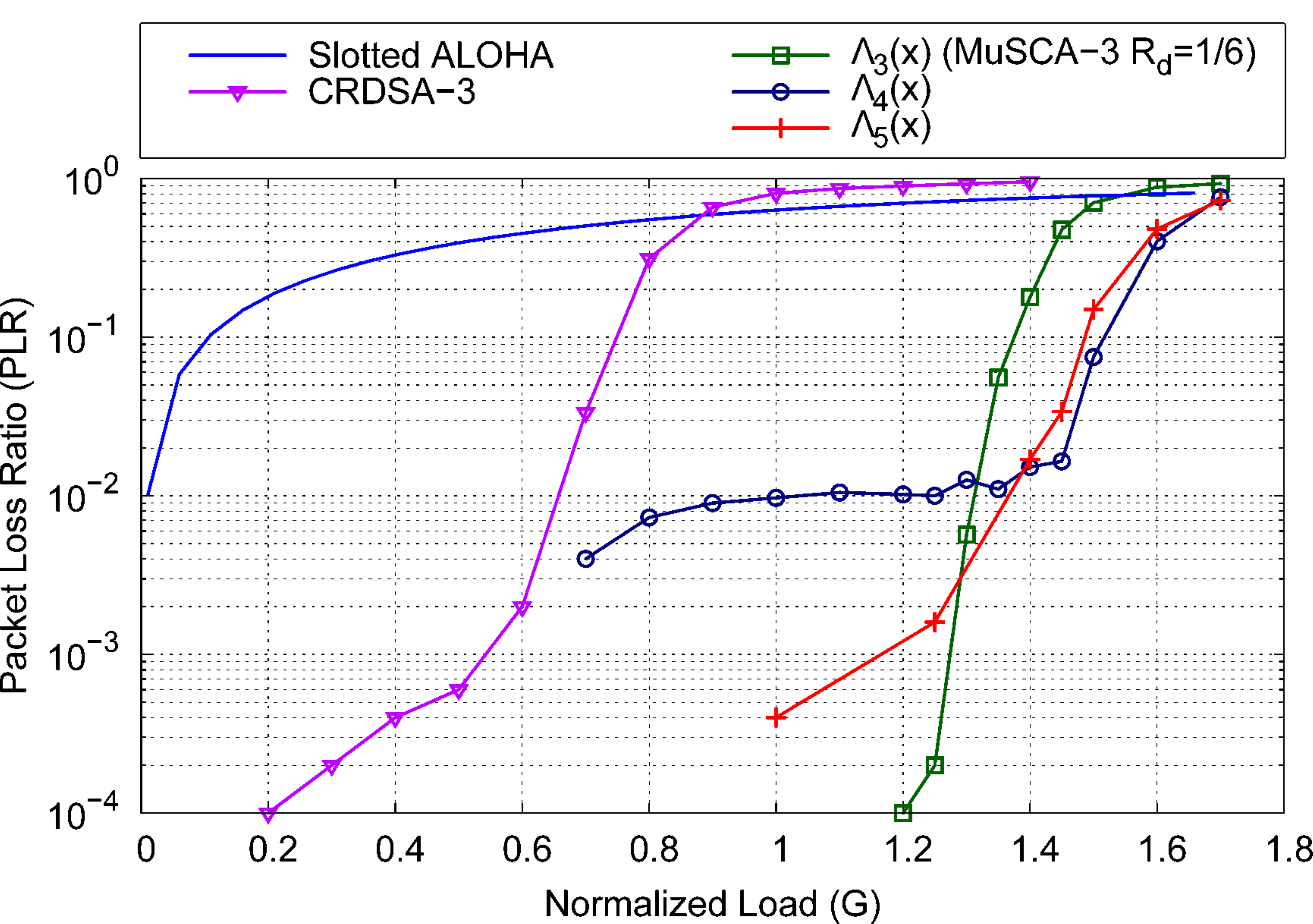}
\label{figPLR}
}
\caption{Simulated results for SA, CRDSA-3, regular MuSCA-3 and irregular degree distributions applied on MuSCA, $E_s/N_0 = 10$ dB}
\label{fig:MuSCAversusCRDSA}
\end{figure}

The optimal distribution $\Lambda_5(x)$ is then studied for several values of SNR. For each SNR, the normalized load $G$ that maximizes the throughput $T$ is identified. Figure~\ref{fig:OptimalDistribution} shows simulation curves for our scheme with the distribution $\Lambda_5(x)$. For SNR higher than 4 dB, the system permits more than 140 users to transmit on a frame of 100 slots. The relation between $T$ and $G$ is almost linear up to $G = 1.4$. That means the probability of successful transmission is maintained close to 1 even if the system is 140\% loaded. At low SNR ($E_s/N_0 = 2$ dB), the high normalized throughput is maintained, $T \approx 1.28$. For a lower level of SNR ($E_s/N_0 = 0$ dB), our scheme still allows up to 76 users.

Figure~\ref{figT} depicts the throughput curves at $E_s/N_0=10$ dB of SA, CRDSA-3 and various probability distributions from Table~\ref{tab:maxT} for MuSCA. We observe that the irregular distribution $\Lambda_5(x)$ containing degrees 1, 2 and 3 achieves a throughput up to 1.43. It provides a throughput gain of 10\% compared to regular MuSCA. The distribution $\Lambda_4(x)$ containing only degrees 2 and 3 obtains a throughput close to the optimal one.

In Figure \ref{figPLR}, we compare the $PLR$ at $E_s/N_0=10$ dB of SA, CRDSA-3 and various irregular degree distributions for MuSCA scheme. We can note that for a normalized load smaller than 1.3, systems with irregular distributions offer higher $PLR$ than MuSCA-3 using turbo code of code rate $R_d = 1/6$. The difference is due to the high $PLR$ of users with degree lower than 3 that take part in irregular distribution. However, for higher value of $G$, the gain in terms of $PLR$ between regular and irregular MuSCA is significant. At $G = 1.4$, $PLR_{irregular} = 0.015$ while $PLR_{MuSCA-3} = 0.179$. The choice of the best distribution must be done according to the target $PLR$. At a $PLR$ lower than $10^{-4}$, regular MuSCA-3 obtains the best results. For a target $PLR = 10^{-2}$, SA operate at extremely low load $G_{SA} \approx 0.01$; CRDSA and CRDSA-3 offer a traffic close to 0.35 and 0.66, respectively \cite{paperCRDSA++} while irregular distributions for MuSCA achieve a traffic close to 1.3.

\begin{figure}[!ht]
\centering
\includegraphics[width=0.9\columnwidth]{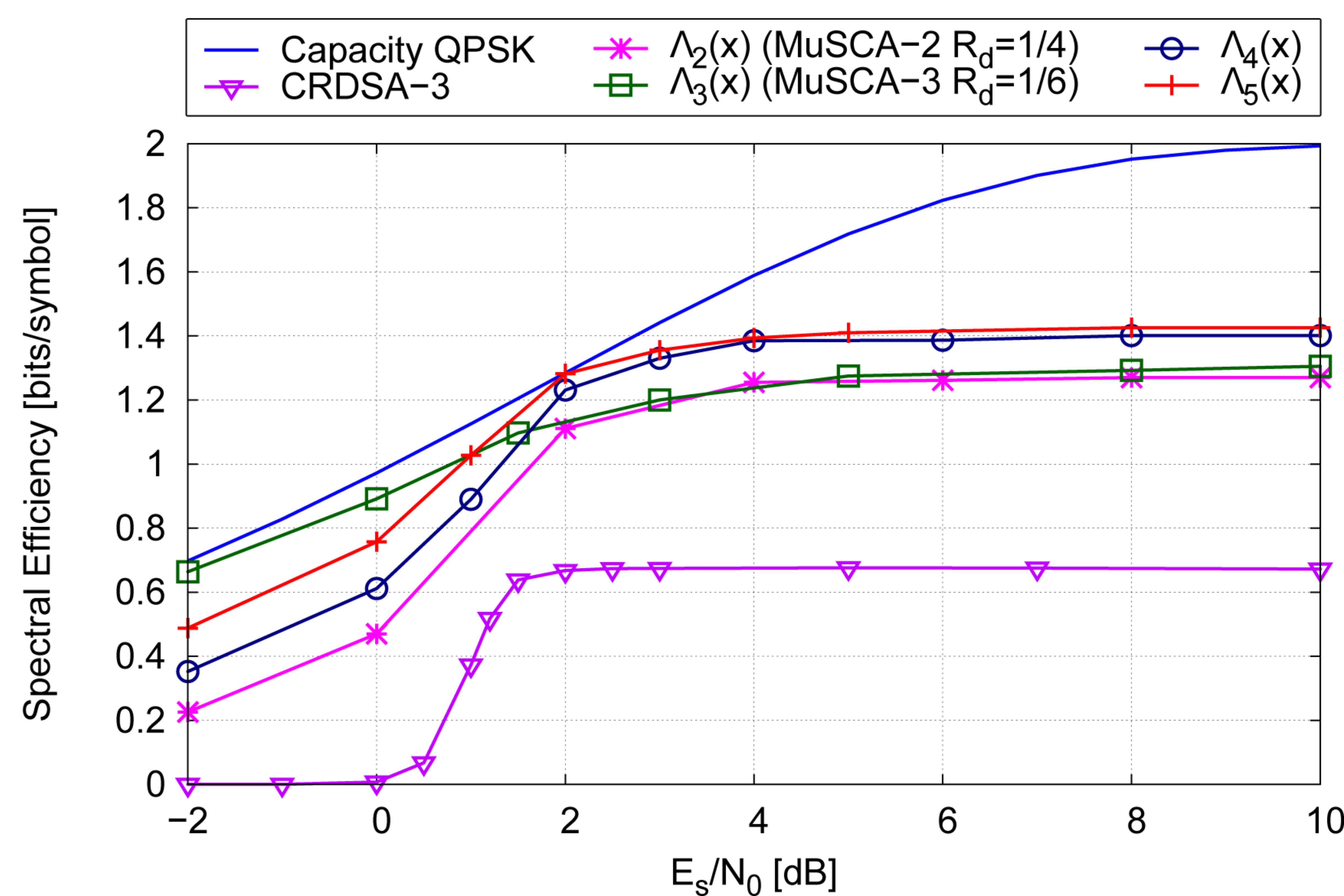}
\caption{Spectral efficiency at various values of SNR} 
\label{fig:Capa}
\end{figure}

To define the best degree distribution for each value of SNR, we quantify the spectral efficiency $S$ as the maximum number of bits per symbol for each distribution. For a given SNR, $S$ is defined by

\begin{equation}
\label{equaC}
S = \frac{max(T)\times k}{L_d},
\end{equation}
where $max(T)$ is the maximum normalized throughput for this SNR and $L_d$ is the number of symbols in a data field. As mentioned in Section \ref{sec:Overview},  $L_d = k/(R_d× N_b× log_2(M))$ symbols, then $S$ can be written as
\begin{equation}
\label{equationC}
S = max(T)\times R_d \times N_b \times log_2(M).
\end{equation}

Figure \ref{fig:Capa} compares numerical results in terms of spectral efficiency from simulations for $\Lambda_2(x)$, $\Lambda_3(x)$, $\Lambda_4(x)$, $\Lambda_5(x)$ and CRDSA-3 scheme with the reference capacity curve of QPSK modulation. At any $E_s/N_0$, we can observe that MuSCA achieves a significant gain compared to CRDSA. At SNR values higher than 1 dB, the distribution $\Lambda_5(x)$ permits to obtain normalized throughput higher than regular MuSCA-3. Particularly, at 2 dB, the spectral efficiency of our scheme is extremely close to the capacity of QPSK modulation. This result is explained by the fact that the sum of 2 QPSK modulated signals can be considered as a signal of a higher order modulation \cite{paperCandH}.

%% file: part5_conclusion.tex
\section{Conclusion and future work}\label{sec:Conclusion}

In this paper, we introduced and analyzed an an improvement of the random access scheme MuSCA. The proposed approach allows users to adopt variable code rates and user degrees, according to a distribution probability. Simulations results show that with the optimal choice of the probability distribution, significant gains in terms of packet loss ratio and normalized throughput are achieved compared to existing random access techniques. For $E_s/N_0 = 10$ dB, our scheme with frames of 100 slots can achieve a normalized throughput close to $1.43$. In future work, we expect to investigate the impact of imperfect channel estimation on the system performance.